\documentclass[12pt]{article}
\usepackage{natbib}
\RequirePackage{etex}
\usepackage[margin=1in]{geometry}  	
\usepackage{amsmath,amssymb,centernot,mathtools}
\usepackage{amsthm}
\usepackage{booktabs}
\usepackage{multirow}
\usepackage{amsmath}
\usepackage{tcolorbox}
\usepackage{float}
\usepackage{caption}
\usepackage{graphicx,subcaption}
\usepackage{adjustbox}
\usepackage{lscape}
\usepackage{adjustbox}
\usepackage{xcolor}
\usepackage{etoolbox}
\usepackage{algorithm}
\usepackage{algorithmic}
\usepackage{tikz}

\usepackage{relsize}
\usepackage{bbm}
\usepackage[sc]{titlesec}
\usepackage[markers,nolists]{endfloat}
\usepackage{enumitem}

\usepackage{authblk}

\newtheorem{theorem}{Theorem}

\newtheorem*{proposition*}{Proposition}

\newtheorem{assumption}{Assumption}
 
 \newtheorem{corollary}{Corollary}

\usepackage{xpatch}
\makeatletter
\xpatchcmd{\proof}{\@addpunct{.}}{\@addpunct{:}}{}{}
\makeatother

\makeatletter
\newcommand{\vast}{\bBigg@{3}}
\newcommand{\Vast}{\bBigg@{4}}
\makeatother

\newcommand\independent{\protect\mathpalette{\protect\independenT}{\perp}}
\def\independenT#1#2{\mathrel{\rlap{$#1#2$}\mkern2mu{#1#2}}}

\makeatletter
\newcommand*{\indep}{%
  \mathbin{%
    \mathpalette{\@indep}{}%
  }%
}
\newcommand*{\nindep}{%
  \mathbin{
    \mathpalette{\@indep}{\not}
  }%
}
\newcommand*{\@indep}[2]{%
  \sbox0{$#1\perp\m@th$}
  \sbox2{$#1=$}
  \sbox4{$#1\vcenter{}$}
  \rlap{\copy0}
  \dimen@=\dimexpr\ht2-\ht4-.2pt\relax
  \kern\dimen@
  {#2}%
  \kern\dimen@
  \copy0 
} 
\makeatother

\usepackage[utf8]{inputenc}
\DeclareUnicodeCharacter{200E}{}
\usepackage{setspace}
\usepackage{url}

\usepackage{xr-hyper}

\usepackage{xcolor}
\definecolor{forestgreen}{RGB}{34,139,34}
\usepackage{hyperref}
\hypersetup{
    colorlinks=true,
    linkcolor=magenta,
    filecolor=magenta, 
    citecolor=forestgreen,      
    urlcolor=blue
}

\titleformat{\section}{\Large\bfseries}{\thesection}{1em}{}
\titleformat{\subsection}{\normalsize\bfseries}{\thesubsection}{1em}{}

\makeatletter
\renewcommand{\boxed}[1]{\text{\fboxsep=.2em\fbox{\m@th$\displaystyle#1$}}}
\makeatother

\makeatletter
\newcommand*{\addFileDependency}[1]{
  \typeout{(#1)}
  \@addtofilelist{#1}
  \IfFileExists{#1}{}{\typeout{No file #1.}}
}
\makeatother

\newcommand*{\myexternaldocument}[1]{%
    \externaldocument{#1}%
    \addFileDependency{#1.tex}%
    \addFileDependency{#1.aux}%
}

\myexternaldocument{webappendix}
\usepackage[stable]{footmisc}

\usepackage{rotating}

\makeatletter
\def\@hangfrom#1{\setbox\@tempboxa\hbox{{#1}}%
      \hangindent 0pt
      \noindent\box\@tempboxa}
\makeatother

\makeatletter
\def\@seccntformat#1{\@ifundefined{#1@cntformat}%
   {\csname the#1\endcsname\quad}  
   {\csname #1@cntformat\endcsname}
}
\let\oldappendix\appendix 
\renewcommand\appendix{%
    \oldappendix
    \newcommand{\section@cntformat}{\appendixname~\thesection\quad}
}
\makeatother

\usepackage[absolute,showboxes]{textpos}

\TPMargin{5pt}

\usepackage{datetime}

\DeclareMathOperator*{\argmin}{arg\,min}



\makeatletter
\g@addto@macro \normalsize {%
 \setlength\abovedisplayskip{5pt plus 0pt minus 0pt}%
 \setlength\belowdisplayskip{5pt plus 0pt minus 0pt}%
}
\makeatother

\begin{document}
\sloppy
\allowdisplaybreaks

\title{ Generalizing conditional average treatment effects from nested randomized trials to all trial-eligible individuals}
\author[1]{Lan Wen}
\author[2]{Issa J. Dahabreh}
\author[3-4]{Yu-Han Chiu}

\affil[1]{Department of Statistics and Actuarial Science, University of Waterloo, Waterloo, ON}
\affil[2]{Department of Epidemiology \& Biostatistics, Harvard T.H. Chan School of Public Health, Boston, MA}
\affil[3]{Department of Epidemiology, Brown University School of Public Health, Providence, RI}
\affil[4]{Center for Gerontology \& Healthcare Research, Brown University School of Public Health, Providence, RI}
\date{}
\maketitle{}

\vspace{-3em}
\begin{abstract}
    \noindent Randomized controlled trials often enroll participants whose characteristics differ from those of a target population, which can limit the generalizability of the estimated treatment effects when effect modifiers differ across populations. While existing generalizability methods primarily focus on estimating the average treatment effect (ATE) in the target population, such summaries may obscure important heterogeneity that is relevant for clinical and policy decision-making. 
    In this work, we illustrate an approach for estimating the conditional average treatment effect (CATE) in a target population of trial-eligible individuals as a function of pre-specified effect modifiers within a nested trial setting. Our approach combines semiparametric theory with flexible estimation: we first estimate nuisance functions using data-adaptive methods and construct pseudo-outcomes from conditional influence functions, then estimate the CATE function via local linear (kernel) regression. Sample splitting and cross-fitting are used to reduce overfitting bias and ensure asymptotic valid inference. 
    Finite-sample performance is assessed via simulations and illustrated in the Coronary Artery Surgery Study (CASS).
    
\end{abstract}

\section{Introduction}
In randomized controlled trials, the primary estimand that is often reported is the average treatment effect (ATE). In many applications, however, investigators are interested not only in the overall treatment effect in the study population, but also in how treatment effects vary across individuals with different baseline characteristics. Such treatment effect heterogeneity is commonly summarized through the conditional average treatment effect (CATE), which characterizes treatment effects as a function of selected effect modifiers \citep{kent2010assessing,kent2016risk,dahabreh2016using,tipton2021beyond,inoue2024machine}. 

At the same time, participants enrolled in trials are often not representative of the broader target population of trial-eligible individuals. When key effect modifiers of the treatment effect differ between the trial and the target population, the treatment effect estimated within the trial may not be applicable to the population in which treatment decisions are ultimately made \citep{cole2010generalizing,stuart2011use,lesko2017generalizing}. 
For instance, the Coronary Artery Surgery Study (CASS), a randomized trial nested within a broader cohort of trial-eligible individuals \citep{dahabreh2019generalizing}, identified prior myocardial infarction and left ventricular function as key effect modifiers of the treatment effect \citep{investigators1984coronary}. In such settings, investigators may wish to generalize treatment effect heterogeneity estimated among trial participants to the broader trial-eligible individuals. 

The literature on generalizability or transportability has primarily focused on extending the ATE from randomized trials to the target population of interest \citep{cole2010generalizing, stuart2011use,westreich2017transportability,lesko2017generalizing,rudolph2017robust,buchanan2018generalizing,dahabreh2019generalizing,dahabreh2020extending,colnet2024causal,ung2025generalizing}, while comparatively less attention has been given to the CATE in this setting \citep{seamans2021generalizability}. Existing approaches for generalizing CATE to a target population have focused on subgroup-specific average treatment effects defined by a small number of discrete (or discretized) covariates, or regression-based methods that rely on parametric specification of nuisance functions and the CATE model \citep{mehrotra2021transporting,robertson2024estimating}. These approaches may become difficult to apply when attempting to 
estimate CATEs as functions of continuous covariates or multiple discrete covariates. In parallel, a growing literature has developed semiparametric and machine learning methods for CATE estimation \citep{abrevaya2015estimating,lee2017doubly,lechner2018modified,kunzel2019metalearners,kennedy2020optimal,semenova2021debiased,nie2021quasi,knaus2021machine,chernozhukov2024conditional,inoue2024machine}. 
However, these methods typically assume that the study population coincides with the target population and therefore do not directly address selective trial participation or generalizability of findings to a broader population.


A key consideration in this setting is distinguishing variables needed to adjust for selective trial participation from variables used to define treatment effect heterogeneity. The former are required to extend inferences from a trial to the target population, while the latter characterize how treatment effects vary within that population. This distinction parallels the separation between confounding adjustment and effect modification in observational studies. In practice, investigators may wish to adjust for a rich set of variables for trial participation while restricting treatment-effect summaries to a smaller set of clinically meaningful effect modifiers,
particularly when certain variables may be considered inappropriate or inequitable for guiding treatment recommendations (e.g., socioeconomic status or insurance status).

In this article, we adapt semiparametric methods for generalizing the CATE from a randomized trial to a target population of trial-eligible individuals under a nested trial design \citep{dahabreh2021study}. The proposed approach integrates recent methodological advances in CATE estimation for randomized trial and observational study settings with semiparametric techniques for generalizability analyses.
Unlike existing work that separately addresses generalizability and flexible estimation of CATE, we develop a unified framework for generalizing CATE  from trial participants to a target population under a nested trial design.

Building on conditional influence function-based methods \citep{chernozhukov2024conditional}, we propose a flexible estimator for the target population CATE. 
In the first stage, we estimate the necessary nuisance components using flexible data-adaptive (machine learning) methods and use these estimates to construct pseudo-outcomes whose conditional expectation corresponds to the target population CATE. In the second stage, we regress the pseudo-outcomes on the selected effect modifiers using local linear (kernel) regression to obtain a smooth approximation of the CATE function. This approach allows continuous and multivariate effect modifiers without requiring discretization or parametric specification of the CATE function.
To ensure valid inference, we use sample splitting and cross-fitting so that the impact of nuisance function estimation on the asymptotic distribution of the final estimator is negligible. We evaluate the finite-sample performance of the methods through simulation studies, and illustrate the method using data from CASS.

\section{Notation and setup}
Consider a trial nested in a cohort of trial-eligible individuals, such as when trial participants are recruited from a broader registry of trial-eligible individuals \citep{dahabreh2021study}.
Let $\mathcal{A}=\{a,a'\}$ denote the set of treatments values evaluated in the randomized trial, and $S$ denote an indicator of trial participation (i.e., $S = 1$ for randomized individuals; $S = 0$ for non-randomized trial-eligible individuals).

Similar to other generalizability analyses, we assume that baseline covariate information ($X$) is available from all individuals in the cohort ($S \in \{0,1\}$), whereas treatment assignment ($A$) and outcome ($Y$) information is only available from randomized trial participants ($S=1$). Treatment and outcome data are not required for non-randomized individuals ($S=0$), as they may be unavailable or deemed less reliable. For simplify exposition, we assume that full adherence to assigned treatment, no measurement error, and no dropout in the trial. Extensions to more complex settings will be considered in future work.

We model the data as independent and identically distributed realizations of the random tuple $O_i=(X_i,S_i,S_i\times A_i,S_i\times Y_i)\sim F$, $i=1,\ldots,n_{s=1}, n_{s=1}+1,\ldots, n_{s=1}+n_{s=0}$, where $n_{s=1}$ denotes the number of randomized individuals, $n_{s=0}$ the number of non-randomized individuals, and $n=n_{s=1}+n_{s=0}$ the total number of trial-eligible individual in the cohort.
For a random variable $Z$, let $\mathbb{P}_n(Z) = {n}^{-1}\sum_{i=1}^n Z_i$, and let $f(z_2)$ and $f(z_2\mid z_1)$ denote the marginal and conditional distributions (density or mass function) of $Z_2$, evaluated at $z_2$, with the latter conditional on $Z_1=z_1$. Henceforth, subscripts $i$ are omitted unless needed.
The causal directed acyclic graphs in Figure A.1 in the Web Appendix A illustrates a possible data generating mechanism for our running example.

Our target of inference is the CATE in the population of trial-eligible individuals, which includes both randomized and non-randomized trial-eligible individuals. Let $V$ denote a set of baseline covariates within $X$ identified a priori as key effect modifiers of interest. For simplicity we focus on $V\in \mathbb{R}$, but the results herein can be generalized to $V\in \mathbb{R}^d$, for $d>1.$
Let $Y^a$ denote the potential outcome under treatment $a \in \mathcal{A}$. For any pair of treatments $a, a' \in \mathcal{A}$, the CATE is defined as $E\bigl(Y^a - Y^{a'} \mid V\bigr),$
which characterizes how treatment effects vary across values of $V$ in the target population.


\section{Generalizing conditional average treatment effects}
Consider the conditional mean potential outcome $E(Y^a \mid V), ~ a \in \mathcal{A}$,
from which the CATE is defined.
We posit the following conditions, which are sufficient for nonparametric identification of the conditional mean potential outcome and the CATE in the target population, and are assumed to hold for each $a\in\mathcal{A}$ \citep{dahabreh2019generalizing}:
(a) \textit{Consistency}: For individuals who receive treatment $A=a$, their observed outcome equals the potential outcome under $a$ (i.e., if $A=a$, then $Y=Y^{a})$. This formulation implicitly assumes that trial participation does not affect the outcome except through treatment (e.g, no trial engagement effect; \citealp{dahabreh2021study}); 
(b) \textit{Conditional exchangeability over treatment $A$ in the trial}: Among randomized individuals ($S=1$), treatment assignment is independent of potential outcomes conditional on baseline covariates $X$, i.e., $Y^{a}\independent A\mid (X,S=1)$; 
(c) \textit{Positivity of treatment assignment in the trial}:  For all covariate values $x$ in the support of $X$ among randomized individuals $S=1$, the probability of assignment to each treatment level is positive, i.e., $P(A = a\mid X = x, S = 1)>0$, $\forall x\in \text{support}(X\mid S=1)$; 
(d) \textit{Conditional exchangeability (generalizability) over trial participation $S$}: Conditional on baseline covariates $X$, the distribution of the potential outcomes is the same across subsets of the target population defined by trial participation, i.e., $Y^{a}\independent S\mid X$; and 
(e) \textit{Positivity of trial participation}: For all covariate values $x$ in the support of $X$ in the target population, the probability of trial participation is positive, i.e., $P(S = 1\mid X = x)>0$, $\forall x\in \text{support}(X)$.

For well-defined treatment interventions in randomized trials, conditions (a)–(c) are generally expected to be satisfied by design. Conditions (d)--(e) allow one to extend inferences about average treatment effects from individuals in the trial to the target population of all trial-eligible individuals. 
The probability of treatment in the trial, $P(A=1\mid X=x,S=1)$, is usually known by design and can therefore be estimated straightforwardly using simple parametric models. In contrast, the trial selection mechanism and outcome models, $P(S=1\mid X)$ and $E(Y\mid A=a,X,S=1)$ are generally unknown and may be complex, motivating the use of data-adaptive methods for their estimation.
Unlike observational studies, where adjustment is needed for confounding between treatment and outcome, here $X$ serves to account for selective participation into the trial.
Under the identifiability conditions listed above, the conditional mean potential outcome under $a$ in the target population of all trial-eligible individuals given $V=v$, $E(Y^{a}\mid V)$, is identified by the following observed data functional:
\begin{align*}
    \psi(a;V) \coloneqq \int_{x} E(Y\mid A=a, X=x, S=1)dF(x\mid V) 
\end{align*}
As such, the CATE, $E(Y^{a=1}\mid V) - E(Y^{a=0}\mid V) $, is identified as:
\begin{align*}
\psi(a=1;V)-\psi(a=0;V) = E[E(Y\mid A=1,X,S=1)\mid V] - E[E(Y\mid A=0,X,S=1)\mid V].
\end{align*} 
For notational simplicity, we define $\theta(V) \coloneqq \psi(a=1;V)-\psi(a=0;V)$. 
In the remainder of the paper, we focus on the estimation of the CATE, identified by the observed data functional $\theta(V)$, for the target population of all trial-eligible individuals.

\section{Conditional influence function for the conditional average treatment effects}
Consider the observed data functional $\mu(a) \coloneqq E[E(Y\mid A=a,X,S=1)]$, which identifies the marginal mean potential outcome $E(Y^a).$
The influence function for $\mu(a)$ \citep{dahabreh2019generalizing},
 can be viewed as the Gateaux derivative of $\mu(a)$ in the direction of a perturbation of the underlying probability distribution $F$ \citep{ichimura2022influence,hines2022demystifying}, which motivates the following conditional influence function.

Now we consider the conditional mean potential outcome under treatment level $a$ (for $a\in \mathcal{A}$), which is identified by $\psi(a;V) = E[ E(Y\mid A=a,X,S=1)\mid V]. $
The corresponding conditional influence function yields conditionally Neyman orthogonal estimating equations for the CATE, and provides a principled basis for constructing pseudo-outcomes for CATE estimation, since it is derived directly from the target estimand \citep{chernozhukov2024conditional}.
Let $F_V$ denote the conditional cumulative distribution function (CDF) of $O$ given $V$, and define the conditional influence function $\phi_V(a;O)$ to be the \textit{Gateaux derivative} of $\psi(a;V) = \psi(a;F_V)$ with respect to $F_V$ at $\epsilon=0$ as the following:
\begin{equation*}
    \frac{d\psi(a;F_{\epsilon,V})}{d\epsilon}\vert_{\epsilon=0} = \int \phi_V(a;O)dH_V(o), ~~~E\left[\phi_V(a;O)\mid V\right]=0,
\end{equation*}
where $H_V$ is a conditional CDF that may be different than $F_V$, and $F_{\epsilon,V} = (1-\epsilon)F_V + \epsilon H_V$ for some $\epsilon\in [0,1].$ 
The Gateaux derivative yields the following expression for the conditional influence function of $\psi(a;V)$, with a proof provided in Web Appendix B.

\begin{theorem}
The conditional influence function of $\psi(a;V)$ is given by:
\begin{align*}
    \phi_V(a;O) &= \Bigg[\frac{S\cdot  I(A=a)}{p(S=1\mid X)p(A\mid X,S=1)}\{Y-\gamma(Z)\} + \{\gamma(a,X,S=1)-\psi(a;V)\}\Bigg].
\end{align*}
where $Z=(A,X,S)$, $\gamma(Z) = E(Y\mid A, X,S)$ and $\gamma(a,X,S=1) = E(Y\mid A=a,X, S=1).$
\label{thm:cif}
\end{theorem}
\begin{corollary}
The conditional influence function of $\theta(V)$ is given by:
\begin{align}
   \phi_{CATE}(O)= & \Bigg[\underbrace{\frac{S\cdot  (2A-1)}{p(S=1\mid X)p(A\mid X,S=1)}}_{\alpha(Z)}\{Y-\underbrace{\gamma(A, X,S)}_{\gamma(Z)}\} + \{\underbrace{\gamma(a=1,X,1)-\gamma(a=0,X,S=1)}_{m(X,S=1,\gamma)}\} - \theta(V)\Bigg]\nonumber
    \\& = \alpha(Z)\{Y-\gamma(Z)\} + m(X,S=1,\gamma) -\theta(V)
    \label{eq:Cifgen}
\end{align}
\label{cor:ciftheta}
where $$\alpha(Z)\coloneqq \dfrac{S\cdot  (2A-1)}{p(S=1\mid X)p(A\mid X,S=1)}; ~~\text{and}~~m(X,S=1,\gamma) = \gamma(a=1,X,S=1)-\gamma(a=0,X,S=1).$$
\end{corollary}

The term $\alpha(Z)$ in \eqref{eq:Cifgen} is the conditional Riesz representer such that $\theta(V) = E[\alpha(Z)\gamma(A,X,S=1)\mid V],$ where functional $\theta(V)=E[m(X,S=1,\gamma)\mid V]$.
This is the conditional analogue of the usual influence function representation for a marginal parameter: just as the marginal expectation of the uncentered influence function (i.e., the influence function plus the parameter) for the ATE recovers the marginal parameter, the conditional expectation of the corresponding uncentered conditional influence function recovers the CATE.
Similar to influence functions for marginal quantities, the conditional influence function \eqref{eq:Cifgen} can be used to construct CATE estimators when data-adaptive methods are used to estimate nuisance functions. Under regularity conditions outlined below, the estimation error in these nuisance functions is asymptotically negligible, so the proposed estimator converges at the same rate as if the nuisance functions were known.
As noted in \cite{chernozhukov2024conditional}, unlike the influence functions for marginal quantities, conditional influence functions cannot be used to facilitate efficiency comparisons between different localization procedures (e.g., kernel smoothing methods, nearest-neighbor or series-based methods) when $V$ is continuous. This limitation arises because the form of localization applied can affect the asymptotic variance. Here, we focus on local linear (kernel) regression for estimating the CATE.

\section{Debiased Machine Learning for the  Conditional Average Treatment Effects}
We adapt a debiased machine learning approach based on local linear regression. In line with prior work (e.g., \citealp{kennedy2023towards,chernozhukov2024conditional}), our results require only mean-square convergence rates for the first-stage nuisance estimators estimated using data-adaptive methods.
To guard against overfitting of the nuisance functions $\alpha(Z)$ and $m(Z)$ as defined above, the algorithm utilizes sample splitting and cross-fitting. 

To estimate $\theta(V)$, we perform localized regression around $V$. 
Following \cite{chernozhukov2024conditional} (see also \citealp{Chernozhukov2018,semenova2021debiased,kennedy2023towards}), we apply sample splitting and cross-fitting to estimate the nuisance functions and the second-stage nonparametric local linear regression. The algorithm is given as follows:
\begin{enumerate}
    \item First, partition the data into $L$ disjoint subsets of equal size ($I_l,$ $l=1,\ldots,L$). For each fold $l=1,\ldots,L$, and for all $i \in I_L$, we estimate $$\hat\xi(O_i;\hat\alpha_{-l}, \hat \gamma_{-l}) = \hat\alpha_{-l}(Z_i)\{Y_i-\hat\gamma_{-l}(Z_i)\} + \hat m_{-l}(X_i,S_i=1,\hat\gamma_{-l}),$$ where the nuisance functions are estimated using $I_{-l}$, i.e., observations that are not in $I_l$. Note that $p(A\mid X,S=1)$ and $\gamma(A,X,S=1)$ are estimated using only individuals with $S=1$ in $I_{-l}$, while $p(S=1\mid X)$ is estimated using all individuals in $I_{-l}.$
    \item $\hat\xi(O_i)\coloneqq \hat\xi(O_i;\hat\alpha_{-l}, \hat \gamma_{-l})$ are then regressed on $V_i$ via locally linear smoother \citep{fan1992design,fan1992variable,fan1994robust,fan1995local}. The locally linear estimator is given by: $\hat\theta(v) = \argmin_{\theta,\delta}\sum_{l=1}^L \sum_{i\in I_l} \left[\left\{\hat \xi(O_i)  - \theta- (V_i-v)^T\delta\right\}^2 K_h(V_i-v)\right].$
\end{enumerate}

\subsection{Asymptotic properties}
Let $K(u)$ be a kernel with $\int K(u)du = 1$ and for bandwidth $h > 0$. Following \cite{chernozhukov2024conditional} (see also \citealp{li2007nonparametric}), we adopt the following conditions on the kernel $K(u)$ and the relevant functionals:
\begin{assumption}[Regularity conditions]
~~
    \begin{enumerate}[label=\Roman*.]
        \item $K$ is a continuous symmetric probability density with support $[-1, 1]$ such $\int K(u)du = 1$ and $K(u)$ and $K(u)u^2$ are bounded;
        \item The bandwidth $h=h_n$ satisfies ${nh}\rightarrow \infty$, $h\rightarrow 0$;
        \item $\lVert \hat \gamma - \gamma\lVert = o_p(h^{1/2})$, $\lVert \hat\alpha - \alpha\lVert = o_p(h^{1/2})$, $\sqrt n\lVert \hat \gamma - \gamma\lVert \lVert \hat\alpha - \alpha\lVert = o_p(h^{1/2}).$
    \end{enumerate}
    \label{ass:reg}
\end{assumption}
Conditions I--II of Assumption \ref{ass:reg} are standard in non-parametric regression. In particular, condition II requires that the ``local sample size'' must increase with sample size while bandwidth $h$ must go to zero \citep{bowman1997applied,li2007nonparametric}.
Condition III plays a central role in the analysis and requires that the mean square product of the nuisance estimators converges to zero at a rate faster than $h^{1/2}n^{-1/2}$, which is slightly stronger than the usual $n^{-1/2}$ product rate condition by an additional factor of $h^{1/2}$. 
\begin{theorem}
    Under the aforementioned conditions (assumption \ref{ass:reg}), $$\sqrt{nh}[\hat\theta(v) - \tilde\theta(v)]\overset{p}{\rightarrow} 0. $$
    \label{thm:two}
\end{theorem}
The proof of Theorem \ref{thm:two} is presented in Web Appendix C. For $V\in \mathbb{R}^d~ (d>1)$, the results follow analogously if (1) assumption 1.I changes to ``$K$ is a continuous, symmetric multivariate probability density on $\mathbb R^d$ with $\int K(u)du = 1$, compact support (e.g., $[-1,1]^d$), and such that $K(u)$ and $||u||^2K(u)$ are bounded"; (2) $nh$ in assumption 1.II changes to $nh^{d}$ \citep{li2007nonparametric}; and (3) $h^{1/2}$ in assumption 1.III changes to $h^{d/2}$ \citep{chernozhukov2024conditional}; then it follows that $\sqrt{nh^d}[\hat\theta(V) - \tilde\theta(V)]\overset{p}{\rightarrow} 0. $

Theorem \ref{thm:two} implies that the estimation errors in the nuisance functions are asymptotically negligible, so the proposed estimator converges at the same rate as if the nuisance functions were known.
Moreover, following \cite{fan1993local} and \cite{fan1994robust}, $E[{\tilde\theta(v)-\theta(v)}^2] = O(1/{nh} + h^4)$, and $\sqrt{nh}(\tilde\theta(v)-\theta(v)-B_h(v))$ converges in distribution to a mean-zero normal random variable with finite variance, where $B_h(v)=\theta''(v)(h^2/2)\int u^2K(u)du$; for $d>1$, replacing $nh$ by $nh^d$ yields the corresponding mean squared error and asymptotic normality results \citep{FanGijbels1996,li2007nonparametric}.

\begin{theorem}
    Under assumption \ref{ass:reg}, and the additional smoothness conditions below:
    \begin{enumerate}[label=\Roman*.]
        \item The functional $\theta(v)$ is twice continuously differentiable with bounded derivatives;
        \item The density $f(v)$ is continuously differentiable and bounded away from zero in a neighborhood of $v$. Moreover, the conditional density of $\xi(O;\alpha,\gamma)$ given $V=v$ exists and is continuous in $(\xi,v)$, where $\xi(O;\alpha,\gamma) = \alpha(Z)\{Y-\gamma(Z)\} + m(X,S=1,\gamma)$;
    \end{enumerate}
    and additionally if $nh^5 = O(1)$, then:
    $$\sqrt{nh}\{\hat\theta(v)- \theta(v)-B_h(v)\} \overset{d}{\longrightarrow}N\left\{0,\frac{\sigma^2(v)\int K(u)^2du}{f(v)}\right\},$$
    where $B_h(v)=\theta''(v)(h^2/2)\int u^2K(u)du,$ and $\sigma^2(v) = Var\{\xi(O;\alpha,\gamma)\mid V=v\}$. 
\end{theorem}
\cite{li2004cross} showed that the cross-validated (CV) selected bandwidth $\hat h$ satisfies $n\hat h^5 = O_p(1).$ Under the conditions of their Theorem 2.1, it follows that $\sqrt{n \hat h}\{\hat\theta(v)- \theta(v)-B_{\hat h}(v)\} \overset{d}{\longrightarrow}N\left\{0,\frac{\sigma^2(v)\int K(u)^2du}{f(v)}\right\}$, where $h$ in $B_h(v)$ is replaced with the CV-selected bandwidth $\hat h$.
As an alternative, one may use a direct plug-in bandwidth selector such as that of \cite{ruppert1995effective}, which targets the theoretically optimal bandwidth but is often observed to oversmooth in practice \citep{loader1999bandwidth}.
For $V\in \mathbb{R}^d~ (d>1)$, the asymptotic normality results can be generalized following Theorem 2.7 of \cite{li2007nonparametric}. 

As in standard nonparametric regression, the estimator is consistent but not exactly centered at $\theta(v)$; instead, it exhibits a bias denoted $B_h(v)$ \citep{wasserman2006all,li2007nonparametric}. Consequently, the estimator is centered at a smoothed version of the effect curve, $\theta^\ast(v) = \theta(v) + B_h(v)$, which, as noted in \cite{wasserman2006all}, complicates the construction of confidence intervals. Following \cite{kennedy2017non}, \cite{wasserman2006all} and related work, we quantify uncertainty in $\hat\theta(v)$ using confidence intervals centered at the smoothed, data-dependent parameter $\theta^\ast(v)$. Define $\beta(v) = (\theta(v),\delta)^T.$ In Web Appendix C, we show that the influence function for $\beta(v)$ can be found to be equal to $$\varphi(O) = \mathbb{D}^{-1}\left[g(V)K_{h}(V-v)\left\{\xi(Z;\eta) - g(V)^T\beta \right\} \right],$$
where $g(V)=(1, V-v)^T$ and $\mathbb{D} = E[g(V)K_{h}(V-v)g(V)^T].$ 

Based on this result, pointwise Wald-style 95\% confidence intervals can be constructed with $\hat\theta(v) \pm 1.96\hat\sigma/\sqrt{n},$ where $\hat\sigma^2$ is the $(1,1)$-th element of sandwich variance estimate $\mathbb{P}_n[\hat\varphi(O)\hat\varphi(O)^T],$ where $\hat\varphi(O)$ defined analogously to $\varphi(O)$, with the nuisance functions replaced by their estimators and $\mathbb{D}$ replaced by $\mathbb{P}_n[g(V)K_h(V-v)g(V)^T].$ 

As such, we should interpret the proposed local linear estimator as a consistent estimator for $\theta^\ast(v)$, which equals a kernel-weighted average of the function $\theta(\cdot)$ in a small neighborhood around $v$, and can be viewed as a smoothed approximation of the true CATE function \citep{wasserman2006all}.
In this sense, for a fixed bandwidth, the estimator targets a local approximation rather than the exact pointwise value at $v$. Nevertheless, as the neighborhood shrinks with increasing sample size, the approximation $\theta^\ast(v)$ converges to $\theta(v)$. Consequently, the estimator still provides a meaningful estimate of the CATE at covariate value $v$.

\section{Simulation studies}
We simulate cohorts of $n=(2500,~5000)$ individuals.
Within each cohort, we vary the number of randomized participants ($n_{s=1}$) across three settings, $n_{s=1} = (125,~500,~1000)$, to assess performance under different proportions of trial participants relative to the target population of trial-eligible individuals.
Following \citet{dahabreh2019generalizing}, we generated data as follows: generation of covariates from a cohort $(X_1,X_2, X_3)$; selection for trial participation ($S$); random treatment assignment among trial participants ($A$); generation of outcome of interest ($Y$), considered under both binary and continuous settings. 
Our interest is in estimating the CATE given by $E(Y^{a=1}-Y^{a=0}\mid X_2)$. Details of the simulation setup, along with detailed results for the continuous case, are provided in Web Appendix D. 

We compare the proposed estimator with a naive alternative. Both estimators use the same first-stage nuisance function estimates and pseudo-outcome construction, but differ in the second-stage regression: the proposed estimator uses local linear regression, whereas the naive estimator fits a simple linear model using $X_2$ as the sole predictor. 
We evaluated the performance of the estimators by comparing them to the true underlying CATE function. 
For the binary outcome, the true underlying CATE function does not admit a closed-form expression and was therefore approximated via Monte Carlo integration of size $n=10^7$. 
For the continuous outcome, the true underlying CATE function is available in closed form and can be evaluated directly. Thus, in this setting, we compare the aforementioned estimators to an oracle estimator that uses the same first-stage nuisance function estimates as the proposed approach, but replaces the second-stage model with the true functional form.

Nuisance functions were estimated using the Super Learner ensemble, which uses cross validation to select the best convex combination of predictions from a library of candidate algorithms \citep{van2007}. The library included generalized linear models and its variants (SL.glm, SL.glm.interaction), gradient boosting (SL.xgboost), generalized additive models with smoothing splines (SL.gam), multivariate adaptive regression Splines (SL.earth), neural networks (SL.nnet) and random forest (SL.ranger). For our proposed estimator, the CATEs were estimated using local linear regression, with the bandwidth parameter selected via cross-validation.
 Estimator performance was summarized using the integrated absolute bias and the root mean squared error (RMSE), with integration performed over the 1000 simulated datasets. We also assessed finite-sample inference by computing integrated coverage probabilities based on the proposed sandwich variance estimator.

\subsection{Simulation results}

\begin{figure}
        \centering
        \includegraphics[width=\linewidth]{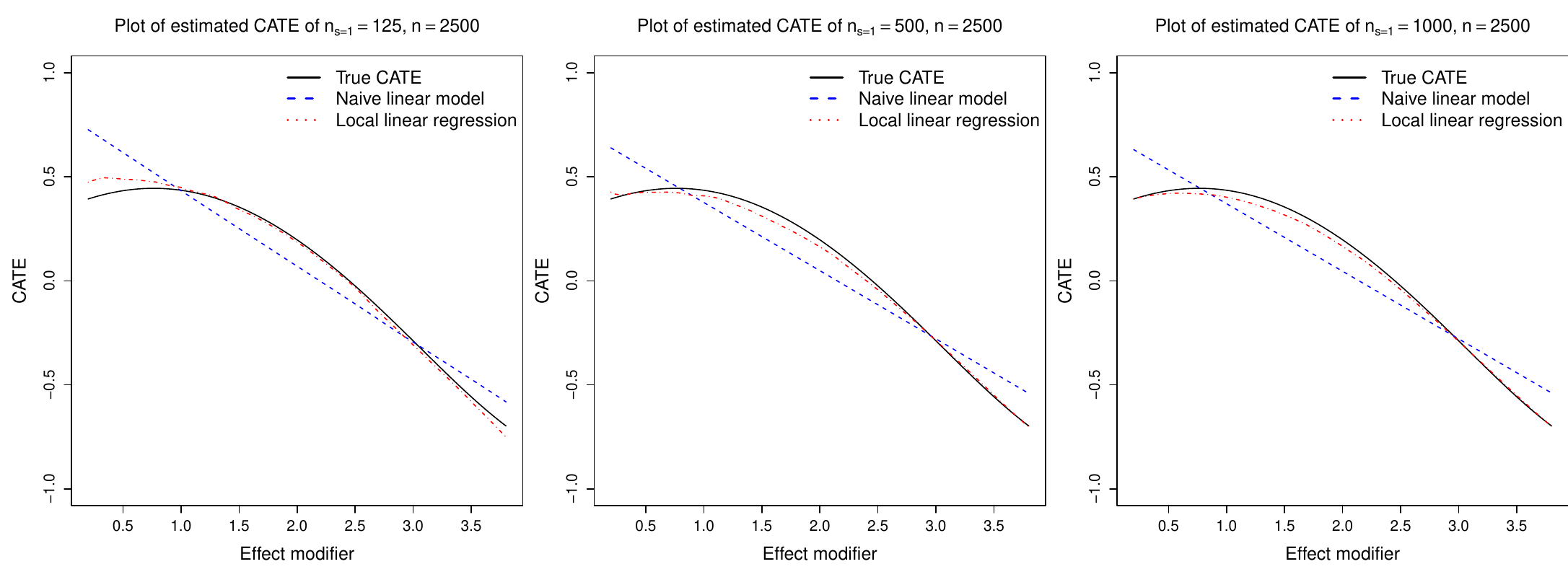}
    \caption{Comparison of proposed estimator with naive estimator for $n=2500$ across trial participation size $n_{s=1}=(125,~500,~1000)$ for \textit{binary outcome}. 95\% coverage probability across values of effect modifier ($X_2$) is shown for the proposed estimator.}
    \label{fig:figurecomparison}
\end{figure}

Simulation results are presented in Table \ref{tab:res2} and Figure \ref{fig:figurecomparison}, with additional results for the continuous outcome provided in Web Appendix D.
For both binary and continuous outcomes, the proposed local linear regression estimator substantially reduced bias relative to the naive regression approach.
Across all sample sizes for the continuous outcome, the oracle linear regression estimator, which correctly specifies the true CATE function, exhibited the smallest integrated absolute bias and achieved coverage probabilities closest to the nominal level. With the exception of the naive regression estimator, the integrated absolute bias of the other estimator(s) decreased as $n_{s=1}$ increased for a fixed cohort size $n$. 

In additional simulation studies reported in Web Appendix D, we compared our proposed estimator with an estimator based only on available trial data. For binary outcomes, the trial-only approach consistently produced biased estimates of the CATE, whereas for continuous outcomes, large bias arose when multiple effect modifiers were present but only a subset was used to estimate the CATE (see Web Appendix D for details).

The confidence intervals for the proposed estimator performed reasonably well, although their coverage probabilities were slightly below the nominal level for continuous outcomes (see Web Appendix D) and closer to nominal for binary outcomes. For instance, when $n_{s=1}=1000$ and $n=2500$, the integrated coverage probability was 92\% for the binary outcome and 88\% for the continuous outcome.
This modest under-coverage may reflect finite-sample smoothing bias in the estimator.
Since inference targets a smoothed version of the CATE rather than the exact underlying curve, the reported intervals are centered on the smoothed parameter, and the remaining bias may still affect coverage in finite samples.
 When evaluating coverage against the average estimated CATE in each simulation setting as an approximation to the smoothed target, we obtain coverage between 93–95\% for $n_{s=1}=(500,1000)$ and between 90–91\% for $n_{s=1}=125$, with similar patterns across both outcome types.

\begin{table}[ht]
\centering
\begin{tabular}{llccc}
\toprule
& Method & Integrated $|\text{Bias}|$ & Integrated RMSE & Integrated CP \\
\midrule
\multicolumn{5}{c}{$n_{s=1}=125,\; n=2500$} \\
 & Local linear regression & 0.022 & 0.345 & 92 \\
 & Linear regression (naive) & 0.096 & 0.322 & 91 \\
\hline
\multicolumn{5}{c}{$n_{s=1}=500,\; n=2500$} \\
 & Local linear regression & 0.020 & 0.167 & 92 \\
 & Linear regression (naive) & 0.101 & 0.168 & 77 \\
\hline
\multicolumn{5}{c}{$n_{s=1}=1000,\; n=2500$} \\
 & Local linear regression & 0.019 & 0.085 & 92 \\
 & Linear regression (naive) & 0.102 & 0.122 & 47 \\
\hline\hline
\multicolumn{5}{c}{$n_{s=1}=125,\; n=5000$} \\
 & Local linear regression & 0.019 & 0.351 & 92 \\
 & Linear regression (naive) & 0.097 & 0.327 & 91 \\
\hline
\multicolumn{5}{c}{$n_{s=1}=500,\; n=5000$} \\
 & Local linear regression & 0.017 & 0.217 & 91\\
 & Linear regression (naive) & 0.102 & 0.187 & 78 \\
\hline
\multicolumn{5}{c}{$n_{s=1}=1000,\; n=5000$} \\
 & Local linear regression & 0.016 & 0.120 & 93 \\
 & Linear regression (naive) & 0.102 & 0.140 & 62 \\
\bottomrule
\end{tabular}
\caption{\label{tab:res2}Integrated Bias, integrated RMSE, and integrated 95\% Coverage Probability (CP) for binary outcome. For the $n_{s=1}=125$ scenarios, the most extreme 5\% of values were trimmed.}
\end{table}

\section{Coronary Artery Surgery Study Data Analysis}
The Coronary Artery Surgery Study (CASS; \citealp{william1983coronary,investigators1984coronary}) was a randomized trial nested within a cohort of trial-eligible patients with stable ischemic heart disease. The trial compared coronary artery bypass grafting plus medical therapy (hereafter ``surgery") against medical therapy alone. Recruitment took place between August 1975 and May 1979, with follow-up for mortality recorded through December 1996. 
The cohort in the data analysis included trial-eligible patients, with a subset randomized into the trial. Moreover, among the randomized group, no censoring was observed during the first ten years of follow-up.

A re-analysis of the CASS data, combining both randomized and observational components, identified effect heterogeneity on the risk difference scale for 10-year mortality across subgroups defined by prior myocardial infarction and impaired left ventricular function, the latter defined as an ejection fraction below 50\% \citep{robertson2021assessing,robertson2023regression}.
We estimate CATEs (risk differences) for 10-year mortality, conditional on baseline ejection fraction, stratified by history of myocardial infarction, using our proposed methodology in the population of trial-eligible individuals. 
Following \cite{robertson2023regression}, our analysis includes 1686 patients, comprising 731 randomized patients (368 assigned to surgery; 363 assigned to medical therapy) and 955 non-randomized patients (430 receiving surgery; 525 receiving medical therapy). Details on the baseline characteristics of the cohort are summarized in \cite{robertson2023regression}. In general, non-randomized patients were more likely to use beta-blockers and had higher left main coronary percent obstruction and left ventricular wall scores.

The nuisance function models adjusted for all covariates considered in \cite{robertson2023regression}. For example, the model for the trial selection mechanism included age, sex, employment status, smoking status, baseline recreational activity level, severity of angina, history of previous myocardial infarction, percent obstruction of the proximal left anterior descending artery, number of diseased vessels, left ventricular wall motion score, systolic blood pressure, diabetes status, ejection fraction, and indicators for regular use of diuretics, nitroglycerin and beta blockers \citep{olschewski1992analysis}.
In the data analysis, nuisance functions were estimated using the Super Learner ensemble with the same library of candidate learners as in simulation studies. The CATE functions, stratified by history of myocardial infarction and evaluated over ejection fraction values ranging from 40\% to 80\%, were estimated using local linear regression with the optimal bandwidth selected via cross-validation.

\begin{figure}
    \centering
    \includegraphics[width=1\linewidth]{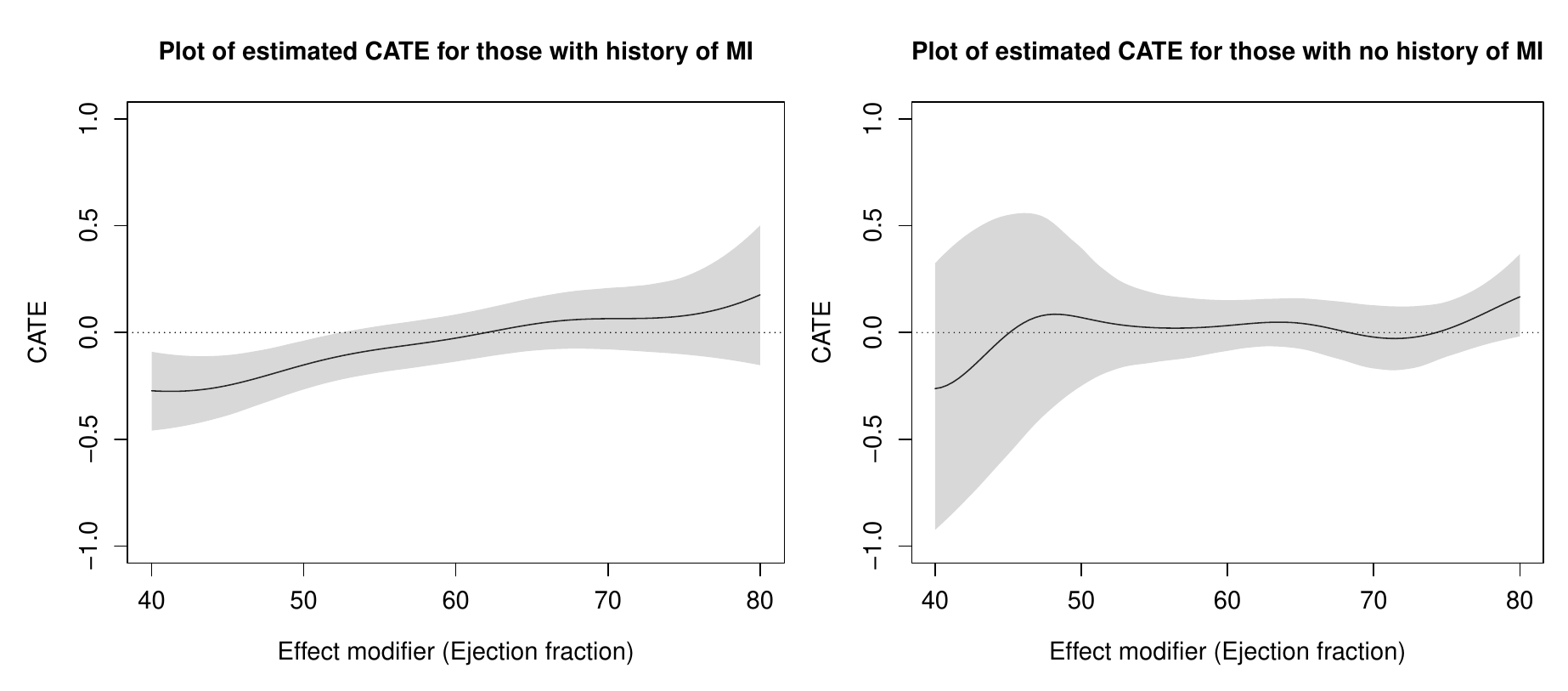}
    \caption{Estimated target population CATE stratified by history of myocardial infarction (MI) over the set of observed ejection fraction values from our proposed estimator}
    \label{fig:dataresults}
\end{figure}

Figure \ref{fig:dataresults} presents the estimated target population CATEs, stratified by history of myocardial infarction, over the observed range of ejection fraction values using our proposed estimator. The results are broadly consistent with prior findings based on the same analysis sample  of \cite{robertson2023regression}, who applied parametric nuisance function models without developing formal semiparametric theory for flexible estimation. Our results suggest that the smoothed CATE functions differ between patients with and without a history of myocardial infarction.

Among patients with a history of myocardial infarction, the smoothed CATE increases approximately linearly from a risk difference of about $-$0.27 at an ejection fraction of 40\% to approximately 0.20 at 80\%. 
In contrast, for patients without a history of myocardial infarction, the CATE initially decreases from around $-$0.26 at 40\%, remains near zero over much of the mid-range (up to roughly 75\%), and then increases to approximately 0.15 at 80\%.
In Web Appendix E, we present a sensitivity analysis comparing the proposed data-integration approach with trial-only estimation of the CATE functions. The results from the trial-only analysis show similar overall trends in the CATE across ejection fraction and MI groups, but the trial-only estimates showed more pronounced fluctuations in the CATE among those with no history of MI compared with the proposed data integration approach.

Using our proposed approach, we estimated a 10-year mortality risk difference of $-0.273$ for individuals with a history of myocardial infarction and an ejection fraction of 40\% (indicative of heart failure; 95\% CI: [$-0.457,~-0.089$]), compared with $-0.152$ at an ejection fraction of 50\% (the lower end of the normal range; 95\% CI: [$-0.266,~-0.038$]). Among individuals without a history of myocardial infarction, the corresponding estimates were $-0.262$ at 40\% ejection fraction (95\% CI: [$-0.871,~0.346$]) and $0.07$ at 50\% ejection fraction (95\% CI: [$-0.249,~0.389$]).
These results suggest that the treatment effect varies with ejection fraction, within levels of myocardial infarction history.

\section{Discussion}
In comparative studies, investigators are often interested in moving beyond the ATE to understand how treatment effects vary across individuals with different baseline characteristics. In practice, this typically involves pre-specifying a subset of effect modifiers based on substantive knowledge or clinical relevance. 
However, even when focusing on subset of well-justified effect modifiers, CATEs estimated using trial data alone may not be directly applied to the broader population of trial-eligible individuals when the distribution of the other effect modifiers differs between the trial and the target population, as is often the case in practice. Accounting for such differences is therefore essential for valid inference on treatment effect heterogeneity and for informing decision-making in the target population. 

Herein, we described a semiparametric framework for estimating target-population CATEs under nested trial sampling. The proposed approach combines data-adaptive estimation of nuisance functions and localized nonparametric regression for continuous effect modifiers. By separating variables used to adjust for selective trial participation from variables used to characterize treatment effect heterogeneity, the framework accommodates settings in which investigators wish to adjust for a rich set of covariates while restricting treatment-effect summaries to a smaller set of clinically meaningful effect modifiers. The resulting approach enables flexible and interpretable estimation of treatment effect heterogeneity in the target population while avoiding restrictive parametric assumptions.

Our inference targets a smoothed version of the CATE, which improves stability in finite samples. Valid inference for the pointwise parameter generally requires either undersmoothing or bias correction. Undersmoothing reduces bias at the cost of infating variance and depends on an ad-hoc bandwidth shrink, whereas bias correction avoids undersmoothing but typically requires estimating additional quantities. For instance, \cite{calonico2018effect}  developed a bias-correction approach that improves finite-sample performance relative to undersmoothing, but requires estimation of higher-order derivatives of the CATE function via additional local polynomial regression.
Future work will investigate how these robust bias-correction techniques can be integrated with the methods developed here to improve coverage accuracy while maintaining stability in finite samples. Additional extensions include accommodating higher-dimensional effect modifiers and developing CATE estimators for non-nested trial settings.



\section*{Acknowledgements}


The authors thank Dr. Sarah Robertson for providing code that enabled replication of the CASS analysis dataset. 
Research reported in this publication was supported by the Natural Sciences and Engineering Research Council of Canada (NSERC) Discovery Grant [RGPIN-2023-03641, DGECR-2023-00455] (Wen); National Institute of General Medical Sciences (NIGMS) award R35GM154888 (Chiu); and National Library of Medicine (NLM) award R01LM013616, National Heart, Lung, and Blood Institute (NHLBI) award R01HL136708, and Patient-Centered Outcomes Research Institute (PCORI) award ME-2021C2-22365 (Dahabreh). The content is solely the responsibility of the authors and does not necessarily represent the official views of the CASS investigators, NIH, NIGMS, NSERC, NLM, NHLBI, PCORI, the PCORI Board of Governors, or the PCORI Methodology Committee. 

\section*{Data availability}
This study used CASS data obtained from the National Heart, Lung, and Blood Institute (NHLBI) Biologic Specimen and Data Repository Information Coordinating Center. 

\newpage
\bibliographystyle{apalike}
\bibliography{refs}

\begin{thebibliography}{}

\bibitem[Abrevaya et~al., 2015]{abrevaya2015estimating}
Abrevaya, J., Hsu, Y.-C., and Lieli, R.~P. (2015).
\newblock Estimating conditional average treatment effects.
\newblock {\em Journal of Business \& Economic Statistics}, 33(4):485--505.

\bibitem[Bowman and Azzalini, 1997]{bowman1997applied}
Bowman, A.~W. and Azzalini, A. (1997).
\newblock {\em Applied smoothing techniques for data analysis: the kernel
  approach with S-Plus illustrations}, volume~18.
\newblock OUP Oxford.

\bibitem[Buchanan et~al., 2018]{buchanan2018generalizing}
Buchanan, A.~L., Hudgens, M.~G., Cole, S.~R., Mollan, K.~R., Sax, P.~E., Daar,
  E.~S., Adimora, A.~A., Eron, J.~J., and Mugavero, M.~J. (2018).
\newblock Generalizing evidence from randomized trials using inverse
  probability of sampling weights.
\newblock {\em Journal of the Royal Statistical Society Series A: Statistics in
  Society}, 181(4):1193--1209.

\bibitem[Calonico et~al., 2018]{calonico2018effect}
Calonico, S., Cattaneo, M.~D., and Farrell, M.~H. (2018).
\newblock On the effect of bias estimation on coverage accuracy in
  nonparametric inference.
\newblock {\em Journal of the American Statistical Association},
  113(522):767--779.

\bibitem[{CASS Principal Investigators}, 1984]{investigators1984coronary}
{CASS Principal Investigators} (1984).
\newblock Coronary artery surgery study (cass): a randomized trial of coronary
  artery bypass surgery: comparability of entry characteristics and survival in
  randomized patients and nonrandomized patients meeting randomization
  criteria.
\newblock {\em Journal of the American College of Cardiology}, 3(1):114--128.

\bibitem[Chernozhukov et~al., 2018]{Chernozhukov2018}
Chernozhukov, V., Chetverikov, D., Demirer, M., Duflo, E., Hansen, C., Newey,
  W., and Robins, J. (2018).
\newblock Double/debiased machine learning for treatment and structural
  parameters.
\newblock {\em The Econometrics Journal}, 21:C1--C68.

\bibitem[Chernozhukov et~al., 2024]{chernozhukov2024conditional}
Chernozhukov, V., Newey, W.~K., and Syrgkanis, V. (2024).
\newblock Conditional influence functions.
\newblock {\em arXiv preprint arXiv:2412.18080}.

\bibitem[Cole and Stuart, 2010]{cole2010generalizing}
Cole, S.~R. and Stuart, E.~A. (2010).
\newblock Generalizing evidence from randomized clinical trials to target
  populations: the actg 320 trial.
\newblock {\em American journal of epidemiology}, 172(1):107--115.

\bibitem[Colnet et~al., 2024]{colnet2024causal}
Colnet, B., Mayer, I., Chen, G., Dieng, A., Li, R., Varoquaux, G., Vert, J.-P.,
  Josse, J., and Yang, S. (2024).
\newblock Causal inference methods for combining randomized trials and
  observational studies: a review.
\newblock {\em Statistical science: a review journal of the Institute of
  Mathematical Statistics}, 39(1):165.

\bibitem[Dahabreh et~al., 2021]{dahabreh2021study}
Dahabreh, I.~J., Haneuse, S. J.~A., Robins, J.~M., Robertson, S.~E., Buchanan,
  A.~L., Stuart, E.~A., and Hern{\'a}n, M.~A. (2021).
\newblock Study designs for extending causal inferences from a randomized trial
  to a target population.
\newblock {\em American journal of epidemiology}, 190(8):1632--1642.

\bibitem[Dahabreh et~al., 2016]{dahabreh2016using}
Dahabreh, I.~J., Hayward, R., and Kent, D.~M. (2016).
\newblock Using group data to treat individuals: understanding heterogeneous
  treatment effects in the age of precision medicine and patient-centred
  evidence.
\newblock {\em International journal of epidemiology}, 45(6):2184--2193.

\bibitem[Dahabreh et~al., 2020]{dahabreh2020extending}
Dahabreh, I.~J., Robertson, S.~E., Steingrimsson, J.~A., Stuart, E.~A., and
  Hernan, M.~A. (2020).
\newblock Extending inferences from a randomized trial to a new target
  population.
\newblock {\em Statistics in medicine}, 39(14):1999--2014.

\bibitem[Dahabreh et~al., 2019]{dahabreh2019generalizing}
Dahabreh, I.~J., Robertson, S.~E., Tchetgen, E.~J., Stuart, E.~A., and
  Hern{\'a}n, M.~A. (2019).
\newblock Generalizing causal inferences from individuals in randomized trials
  to all trial-eligible individuals.
\newblock {\em Biometrics}, 75(2):685--694.

\bibitem[Fan, 1992]{fan1992design}
Fan, J. (1992).
\newblock Design-adaptive nonparametric regression.
\newblock {\em Journal of the American statistical Association},
  87(420):998--1004.

\bibitem[Fan, 1993]{fan1993local}
Fan, J. (1993).
\newblock Local linear regression smoothers and their minimax efficiencies.
\newblock {\em The annals of Statistics}, pages 196--216.

\bibitem[Fan and Gijbels, 1992]{fan1992variable}
Fan, J. and Gijbels, I. (1992).
\newblock Variable bandwidth and local linear regression smoothers.
\newblock {\em The Annals of Statistics}, pages 2008--2036.

\bibitem[Fan and Gijbels, 1996]{FanGijbels1996}
Fan, J. and Gijbels, I. (1996).
\newblock {\em Local Polynomial Modelling and Its Applications}.
\newblock Chapman \& Hall, London.

\bibitem[Fan et~al., 1995]{fan1995local}
Fan, J., Heckman, N.~E., and Wand, M.~P. (1995).
\newblock Local polynomial kernel regression for generalized linear models and
  quasi-likelihood functions.
\newblock {\em Journal of the American Statistical Association},
  90(429):141--150.

\bibitem[Fan et~al., 1994]{fan1994robust}
Fan, J., Hu, T.-C., and Truong, Y.~K. (1994).
\newblock Robust non-parametric function estimation.
\newblock {\em Scandinavian journal of statistics}, pages 433--446.

\bibitem[Hines et~al., 2022]{hines2022demystifying}
Hines, O., Dukes, O., Diaz-Ordaz, K., and Vansteelandt, S. (2022).
\newblock Demystifying statistical learning based on efficient influence
  functions.
\newblock {\em The American Statistician}, 76(3):292--304.

\bibitem[Ichimura and Newey, 2022]{ichimura2022influence}
Ichimura, H. and Newey, W.~K. (2022).
\newblock The influence function of semiparametric estimators.
\newblock {\em Quantitative Economics}, 13(1):29--61.

\bibitem[Inoue et~al., 2024]{inoue2024machine}
Inoue, K., Adomi, M., Efthimiou, O., Komura, T., Omae, K., Onishi, A.,
  Tsutsumi, Y., Fujii, T., Kondo, N., and Furukawa, T.~A. (2024).
\newblock Machine learning approaches to evaluate heterogeneous treatment
  effects in randomized controlled trials: a scoping review.
\newblock {\em Journal of Clinical Epidemiology}, 176:111538.

\bibitem[Kennedy, 2023]{kennedy2023towards}
Kennedy, E.~H. (2023).
\newblock Towards optimal doubly robust estimation of heterogeneous causal
  effects.
\newblock {\em Electronic Journal of Statistics}, 17(2):3008--3049.

\bibitem[Kennedy et~al., 2020]{kennedy2020optimal}
Kennedy, E.~H. et~al. (2020).
\newblock Optimal doubly robust estimation of heterogeneous causal effects.
\newblock {\em arXiv preprint arXiv:2004.14497}, 5.

\bibitem[Kennedy et~al., 2017]{kennedy2017non}
Kennedy, E.~H., Ma, Z., McHugh, M.~D., and Small, D.~S. (2017).
\newblock Non-parametric methods for doubly robust estimation of continuous
  treatment effects.
\newblock {\em Journal of the Royal Statistical Society Series B: Statistical
  Methodology}, 79(4):1229--1245.

\bibitem[Kent et~al., 2016]{kent2016risk}
Kent, D.~M., Nelson, J., Dahabreh, I.~J., Rothwell, P.~M., Altman, D.~G., and
  Hayward, R.~A. (2016).
\newblock Risk and treatment effect heterogeneity: re-analysis of individual
  participant data from 32 large clinical trials.
\newblock {\em International journal of epidemiology}, 45(6):2075--2088.

\bibitem[Kent et~al., 2010]{kent2010assessing}
Kent, D.~M., Rothwell, P.~M., Ioannidis, J.~P., Altman, D.~G., and Hayward,
  R.~A. (2010).
\newblock Assessing and reporting heterogeneity in treatment effects in
  clinical trials: a proposal.
\newblock {\em Trials}, 11(1):85.

\bibitem[Knaus et~al., 2021]{knaus2021machine}
Knaus, M.~C., Lechner, M., and Strittmatter, A. (2021).
\newblock Machine learning estimation of heterogeneous causal effects:
  Empirical monte carlo evidence.
\newblock {\em The Econometrics Journal}, 24(1):134--161.

\bibitem[K{\"u}nzel et~al., 2019]{kunzel2019metalearners}
K{\"u}nzel, S.~R., Sekhon, J.~S., Bickel, P.~J., and Yu, B. (2019).
\newblock Metalearners for estimating heterogeneous treatment effects using
  machine learning.
\newblock {\em Proceedings of the national academy of sciences},
  116(10):4156--4165.

\bibitem[Lechner, 2018]{lechner2018modified}
Lechner, M. (2018).
\newblock Modified causal forests for estimating heterogeneous causal effects.
\newblock {\em arXiv preprint arXiv:1812.09487}.

\bibitem[Lee et~al., 2017]{lee2017doubly}
Lee, S., Okui, R., and Whang, Y.-J. (2017).
\newblock Doubly robust uniform confidence band for the conditional average
  treatment effect function.
\newblock {\em Journal of Applied Econometrics}, 32(7):1207--1225.

\bibitem[Lesko et~al., 2017]{lesko2017generalizing}
Lesko, C.~R., Buchanan, A.~L., Westreich, D., Edwards, J.~K., Hudgens, M.~G.,
  and Cole, S.~R. (2017).
\newblock Generalizing study results: a potential outcomes perspective.
\newblock {\em Epidemiology}, 28(4):553--561.

\bibitem[Li and Racine, 2004]{li2004cross}
Li, Q. and Racine, J. (2004).
\newblock Cross-validated local linear nonparametric regression.
\newblock {\em Statistica Sinica}, pages 485--512.

\bibitem[Li and Racine, 2007]{li2007nonparametric}
Li, Q. and Racine, J.~S. (2007).
\newblock {\em Nonparametric econometrics: theory and practice}.
\newblock Princeton University Press.

\bibitem[Loader, 1999]{loader1999bandwidth}
Loader, C.~R. (1999).
\newblock Bandwidth selection: classical or plug-in?
\newblock {\em The Annals of Statistics}, 27(2):415--438.

\bibitem[Mehrotra et~al., 2021]{mehrotra2021transporting}
Mehrotra, M.~L., Westreich, D., Glymour, M.~M., Geng, E., and Glidden, D.~V.
  (2021).
\newblock Transporting subgroup analyses of randomized controlled trials for
  planning implementation of new interventions.
\newblock {\em American journal of epidemiology}, 190(8):1671--1680.

\bibitem[Nie and Wager, 2021]{nie2021quasi}
Nie, X. and Wager, S. (2021).
\newblock Quasi-oracle estimation of heterogeneous treatment effects.
\newblock {\em Biometrika}, 108(2):299--319.

\bibitem[Olschewski et~al., 1992]{olschewski1992analysis}
Olschewski, M., Schumacher, M., and Davis, K.~B. (1992).
\newblock Analysis of randomized and nonrandomized patients in clinical trials
  using the comprehensive cohort follow-up study design.
\newblock {\em Controlled clinical trials}, 13(3):226--239.

\bibitem[Robertson et~al., 2021]{robertson2021assessing}
Robertson, S.~E., Leith, A., Schmid, C.~H., and Dahabreh, I.~J. (2021).
\newblock Assessing heterogeneity of treatment effects in observational
  studies.
\newblock {\em American Journal of Epidemiology}, 190(6):1088--1100.

\bibitem[Robertson et~al., 2023]{robertson2023regression}
Robertson, S.~E., Steingrimsson, J.~A., and Dahabreh, I.~J. (2023).
\newblock Regression-based estimation of heterogeneous treatment effects when
  extending inferences from a randomized trial to a target population.
\newblock {\em European journal of epidemiology}, 38(2):123--133.

\bibitem[Robertson et~al., 2024]{robertson2024estimating}
Robertson, S.~E., Steingrimsson, J.~A., Joyce, N.~R., Stuart, E.~A., and
  Dahabreh, I.~J. (2024).
\newblock Estimating subgroup effects in generalizability and transportability
  analyses.
\newblock {\em American journal of epidemiology}, 193(1):149--158.

\bibitem[Rudolph and Laan, 2017]{rudolph2017robust}
Rudolph, K.~E. and Laan, M.~J. (2017).
\newblock Robust estimation of encouragement design intervention effects
  transported across sites.
\newblock {\em Journal of the Royal Statistical Society Series B: Statistical
  Methodology}, 79(5):1509--1525.

\bibitem[Ruppert et~al., 1995]{ruppert1995effective}
Ruppert, D., Sheather, S.~J., and Wand, M.~P. (1995).
\newblock An effective bandwidth selector for local least squares regression.
\newblock {\em Journal of the American Statistical Association},
  90(432):1257--1270.

\bibitem[Seamans et~al., 2021]{seamans2021generalizability}
Seamans, M.~J., Hong, H., Ackerman, B., Schmid, I., and Stuart, E.~A. (2021).
\newblock Generalizability of subgroup effects.
\newblock {\em Epidemiology}, 32(3):389--392.

\bibitem[Semenova and Chernozhukov, 2021]{semenova2021debiased}
Semenova, V. and Chernozhukov, V. (2021).
\newblock Debiased machine learning of conditional average treatment effects
  and other causal functions.
\newblock {\em The Econometrics Journal}, 24(2):264--289.

\bibitem[Stuart et~al., 2011]{stuart2011use}
Stuart, E.~A., Cole, S.~R., Bradshaw, C.~P., and Leaf, P.~J. (2011).
\newblock The use of propensity scores to assess the generalizability of
  results from randomized trials.
\newblock {\em Journal of the Royal Statistical Society Series A: Statistics in
  Society}, 174(2):369--386.

\bibitem[Tipton, 2021]{tipton2021beyond}
Tipton, E. (2021).
\newblock Beyond generalization of the ate: designing randomized trials to
  understand treatment effect heterogeneity.
\newblock {\em Journal of the Royal Statistical Society Series A: Statistics in
  Society}, 184(2):504--521.

\bibitem[Ung et~al., 2025]{ung2025generalizing}
Ung, L., VanderWeele, T.~J., and Dahabreh, I.~J. (2025).
\newblock Generalizing and transporting causal inferences from randomized
  trials in the presence of trial engagement effects.
\newblock {\em Epidemiology}, 36(4):500--510.

\bibitem[van Der~laan et~al., 2007]{van2007}
van Der~laan, M.~J., Polley, E.~C., and Hubbard, A.~E. (2007).
\newblock Super learner.
\newblock {\em Statistical applications in genetics and molecular biology}, 6.

\bibitem[Wasserman, 2006]{wasserman2006all}
Wasserman, L. (2006).
\newblock {\em All of nonparametric statistics}.
\newblock Springer.

\bibitem[Westreich et~al., 2017]{westreich2017transportability}
Westreich, D., Edwards, J.~K., Lesko, C.~R., Stuart, E., and Cole, S.~R.
  (2017).
\newblock Transportability of trial results using inverse odds of sampling
  weights.
\newblock {\em American journal of epidemiology}, 186(8):1010--1014.

\bibitem[William et~al., 1983]{william1983coronary}
William, J., Russell, R., Nicholas, T., et~al. (1983).
\newblock Coronary artery surgery study (cass): a randomized trial of coronary
  artery bypass surgery.
\newblock {\em Circulation}, 68(5):939--950.

\end{thebibliography}

\end{document}